\documentclass[twocolumn]{emulateapj}
\submitted{Science with Parkes @ 50 Years Young, 31 Oct. -- 4 Nov., 2011}

\shorttitle{Understanding our Galaxy - 50 years at Parkes}
\shortauthors{J.L. Caswell}

\begin{document}


\title{Understanding our Galaxy - key contributions\\
     from the Parkes telescope.}


\author{J. L. Caswell}
\affil{CSIRO Astronomy and Space Science, ATNF}
\email{james.caswell@csiro.au}


\begin{abstract}
Young massive stars, with their spectacular masers and HII regions, 
dominate our Galaxy, and are a cornerstone for understanding Galactic 
structure.  I will highlight the role of Parkes in contributing to 
these studies  - past, present and future.

\end{abstract}





\section{Introduction}

On the first day of our symposium. the contributions were entertaining 
and knowledgable reflections on the early years at Parkes - chiefly 
recalling an era from which there are few pioneers left with 
personal experience.   
Subsequent days focus on active work still continuing at Parkes.  I am 
pleased to be the bridge from the past to the present, especially to 
the next session of the symposium, dealing with current studies of young  
massive stars and their masers.  
  
My main theme will be to show how the Parkes studies of masers and 
related objects contribute to revolutionizing the picture of our 
Galaxy, its content and its structure.   

\section{Galactic structure}




\subsection{An early view}

Radio astronomy, with its ability to see through the dusty disc of our 
Galaxy, gave us the first quantitative realization of the likely extent 
of the Galactic disc, and hints of its full spiral structure.  The 
history of these discoveries stemming from the earliest HI observations 
is beautifully summarized by Oort, Kerr \& Westerhout (1958).  They 
present an adventurous first attempt at revealing spiral structure,  
tempered by an acknowledgment of its very preliminary nature.  They also 
include prescient musings on the likely role of large-scale interstellar 
magnetic fields influencing the appearance of the spiral arms.  

This early picture of spiral structure was largely reinforced by 
subsequent improved HI observations, and by the newer tool of CO 
observations, which map the dense molecular clouds that are somewhat 
more closely related to the expected massive star distribution.


\subsection{Tracers of massive stars, the dominant 
component of spiral structure}

While the spiral structure of our Galaxy depicted by HI gas was of great 
interest, the spiral structure seen photographically in nearby 
galaxies is traced chiefly by a very different population - 
the hot, young, massive blue stars, and the HII regions that surround 
them.  Optical study of these objects in our Galaxy is impeded by 
dust and obscuration, so we seek the radio equivalent;  the HII regions, 
in particular, are detectable from both their radio continuum emission 
and their recombination line emission (with the corresponding capability 
of measuring systemic velocities).

\subsection{The value of Parkes for Galactic structure}

Observations between 1.4 and 10 GHz are excellent for the HII  
measurements, and thus readily pursued at Parkes.  This is complemented 
by another key attribute of Parkes - its southern hemisphere location 
(with the Galactic Centre passing nearly overhead) and thus accessing 
the 70 per cent of Galactic disc that is most important for study of its  
structure.  

The systemic velocity of an HII region, measured from its radio 
recombination lines, is an excellent first start at determining its 
distance, able to yield a `kinematic' distance based on a simple 
rotation curve for the Galaxy.  However, the kinematic distances for 
Galactic locations nearer the Centre than the sun are generally 
ambiguous, and the ambiguity needs to be resolved by an alternative 
means.  Fortunately, an enhancement to the Parkes dish, the Parkes 
interferometer, provided a useful discriminator and  was being developed 
at just the right time.

\subsection{The Parkes HI interferometer}

Yesterday, Ron Ekers described the innovative design of the Parkes 
interferometer and here I will remind you of its modification and novel 
use by Radhakrishnan to study the HI spectra towards strong continuum 
sources (Radhakrishnan et al. 1972a).  The primary purpose was to 
measure the properties of absorbing HI clouds;  but, as a by-product, 
for sources in the Galactic plane, it could determine kinematic 
distances of Galactic sources, as amply demonstrated by Radhakrishnan 
and Miller Goss (Radhakrishnan et al. 1972b).
My first years at Radiophysics overlapped this work, shortly before Rad 
and Miller both departed from Australia.  I was excited at the prospect 
of getting more distances - a project that would be abandoned if I did 
not adopt it.   
I was blessed to inherit a talented team of engineers (John Murray, 
Dave Cooke and Doug Cole), complemented by astronomy advice and help 
from Rob Roger, visiting at that time from Penticton (where we had 
previously worked together), and fresh from planning an HI 
interferometer there.  Upgrades to the Parkes interferometer 
allowed us to achieve excellent results, leading to a large number of 
distance determinations for Galactic radio sources, both supernova 
remants and HII regions (Caswell et al. 1975).

\section{New surveys of the Galactic disc}

Progress in mapping the Galaxy had been good, but needed the impetus of 
new surveys of Galactic radio objects.  
We chose to survey the Galactic plane at 5 GHz in another ambitious 
large project.  This was conducted with Raymond Haynes, firstly in the 
continuum (Haynes, Caswell \& Simons 1978), and then with a 
recombination line follow-up (Caswell \& Haynes 1987).
This greatly improved our assessment of spiral arms in the southern sky, 
especially the Carina arm.

\subsection{OH and water masers - the early years}

Until now, I have kept to a chronological order, but that eventually 
becomes impossible in the  real world of overlapping events.  
It seems appropriate to segue into a quote from Dean Kamen: 
`People take the longest possible paths, digress to numerous dead ends, 
and make all kinds of mistakes.  Then historians come along and write 
summaries of this messy, non-linear process and make it appear like a 
simple straight line'.  More succinct is a related sentiment from Mark 
Twain:  
`In the real world nothing happens at the right place 
and the right time.  It is the job of journalists and historians to 
correct that'.  Hindsight, even if unintentional,  modifies the story.  
But, overall,  in my presentation of the subsequent research, I will try 
to retain the logic and motivation that drove it.  

I now backtrack a few years to show where masers fit into the picture.  
OH maser research in the early 1970s was taking another step 
forward, with searches for new masers planned.  I was fortunate to 
be able to join Brian Robinson and Miller Goss in these developments.

How many varieties of OH masers?   The sample known so far was now 
growing to the point of recognising different varieties, one of which  
was a large population associated with massive star formation (Robinson, 
Caswell \& Goss 1974).

OH masers were certainly fascinating; surely they would also be useful?
To answer this question, our intention was to conduct OH survey projects 
in a prompt and orderly manner but were then interrupted by  `an 
opportunity that was too good to refuse' - an opportunity 
to observe water masers in the southern sky.  

Ken Johnston from the Naval Research Laboratory had a receiver, we had 
OH targets, and a first round of improvements to the dish surface at 
Parkes made it a viable instrument at 22 GHz.  The sensitivity 
was adequate for detecting strong maser emission, and the major   
challenge was a small beamsize above 20 GHz - a valuable 
property but with associated problems from pointing 
errors and uncertainties in our target positions - a hexagonal grid 
search was needed every time (Johnston et al. 1972).  Calibration was 
also a challenge.  Ken Kellerman reminded us yesterday of early planet 
observations at Parkes, noting that these were not subsequently 
continued.  In fact, we found Jupiter to be the solution to our 
calibration problem, since at that epoch it was a southerly object near 
declination -20 degrees.  

For several years we then conducted unbiased surveys for OH masers in 
the Galactic disc (e.g. Caswell, Haynes \& Goss 1980), with follow-up of 
water masers (Batchelor et al. 1980).  Coincidence of water 
masers with OH was a matter of dispute - how closely associated were 
they?  Our position accuracy at Parkes was limited to about 10 arcsec, 
at which level the coincidence seemed good.  To understand their 
relationship we needed more 
precise positions.  Some of our masers were within reach of the VLA, and 
Rick Forster convinced me that it could provide a partial solution to 
our problem.  

It was an ambitious project for the VLA when it was scheduled in 1983.
For a while, Rick and I had the VLA record for data processing 
requirements:  our experiment lasted only a few days, but we had 
spectral line data at long baselines for nearly 100 sources, and needed 
high resolution spectral line cubes.  The results amply 
repaid our efforts, and demonstrated the extremely close 
association for many pairs, plus the excess of water maser sites 
compared to OH (Forster \& Caswell 1989).  

Eventually it was possible to obtain maser positions in the true 
southern 
sky, using the Compact Array of the AT.  There followed a  
productive combination of Parkes spectra and  Compact Array positions 
for OH masers, but it was another decade before we could extend the 
ATCA  studies to the 22-GHz water masers.

\subsection{The methanol `explosions'}

The discovery of maser emission at 12 GHz was the first of two major 
impacts by methanol on star formation maser studies.  
In an abrupt diversion of effort, with an `off-the-shelf' commercial 
receiver spedily made suitable for Parkes by Kel 
Wellington, we were able to use our newly-determined OH maser 
positions to conduct a search for further examples of the newly 
discovered 12-GHz methanol transition -  with considerable success 
(Norris et al. 1987).

Scarcely had we returned to pick up the threads of the OH and water 
research when the next methanol impact was upon us - the discovery 
in 1991 (Menten 1991) of methanol maser emission at 6668 MHz - which 
was found to usually surpass both OH and 12-GHz methanol in intensity. 
Once again the agility of Parkes, and our receiver group, was 
demonstrated, as an excellent new receiver was rapidly deployed and 
enabled us to reveal the full importance of this transition (Caswell et 
al. 1995).  We finished up with several hundred methanol masers, 
allowing us to explore the common properties, unusual properties, and 
even characterise the typical variability, with the first hints of the 
exciting possibility that some might be periodic variables (Caswell, 
Vaile \& Ellingsen 1995).  Periodicity was eventually confirmed in South 
Africa several years later (Goedhart et al. 2004, 2009).
John Whiteoak was a key member of our team and recognised the role that 
the Compact array could play in methanol studies, despite 6.6 GHz being 
a frequency outside the nominal ATCA coverage.  ATCA accurate positions 
of the OH  and methanol masers then established unequivocally their 
intimate association (Caswell, Vaile \& Forster 1995), with their common 
source of excitation provided by a massive embryonic central massive 
star.

\section{More surveys, and their purpose}

Could those searches for methanol masers be described as a survey?
Yesterday, we heard several views on the purposes of surveys, prompted 
by a presentation from Jasper Wall and Carole Jackson.  I have been 
engaged in rather a lot of surveys, necessitated by exploring the 
Galaxy.  These are some of my reflections:

The purpose obviously depends on previous knowledge, and the space 
density of the expected population.

If very few objects are known, then the main purpose may be simply to 
expand the sample.  

If the space density is intrinsically low, much of the importance of a 
uniformly sensitive large survey is to define the regions where there is 
nothing!

In a few cases, e.g. a finite Galactic population, we may discover the 
whole population!

The last point is especially exciting, and seemed applicable to the 
methanol masers.  

\subsection{The MMB survey}

To fully exploit the value of 6.6-GHz methanol masers, we  needed a 
'proper' survey that was sensitive,  had no bias to pre-selected 
targets, and covered the quite large area of the whole Galactic disc.  
We questioned whether to use Parkes (necessarily requiring a multi-beam 
receiver to map with adequate speed and sensitivity) or the ATCA?  We 
chose the best of both options, with Parkes for the survey proper, and 
the ATCA as an integral part of the 
project in providing precise positions to arcsecond accuracy.

The Methanol Multibeam (MMB) project proposal to build the receiver was 
submitted 2001 February, early in the year of the Parkes 40th birthday, 
with Mal Sinclair as project leader and J. Caswell as project scientist.

It was a collaborative venture with Jodrell Bank, with Jim Cohen leading 
the UK part.   Receiver construction was shared between our two 
institutions and was ready for testing on the Parkes telescope 2006 January.  

Jim Cohen and I had planned the survey strategy and worked intensively 
together to get the survey running smoothly and productively immediately 
after `first light' on Australia Day of 2006.  It was a memorable first 
year of observations, with Jim Cohen present at 
all our sessions, and Jim's wife Pat welcomed as an additional 
enthusiastic team member when needed.  

Sadly, to our great dismay, Jim died late in 2006, exactly 5 years ago.   
Over the following few years, we have achieved our goal of completing 
the survey, and Pat Cohen has been delighted to see these outcomes from 
the work that consumed so much of Jim's time in the final year of his 
life;  the legacy of the survey is a fitting tribute to his efforts 
(Green et al. 2009;  Caswell et al. 2010).  

The outcomes from the survey will be extensive, with many productive 
follow-ups already completed, and others continuing.   There are two 
areas in particular where the masers have high impact:   firstly the   
far-reaching implications for Galactic structure, where considerable 
progress has already been made, and which we will return to later;  and 
secondly, the full characterisation of each maser site, by its mass, 
evolutionary state and environment, and thereby contributing 
to the ongoing investigation of the poorly understood mechanism of high 
mass star formation,

\section{Excited-state OH transitions at Parkes}

Once again we step back, this time to catch up on the OH maser progress.  
Although OH maser studies of star formation regions mostly focus on 
the ground-state transitions at 1665 and 1667 MHz, the 6035 and 6030-MHz 
excited-state transitions have recently been 
recognised as equally valuable, and sometimes more so.  For many years, 
very few telescopes were equipped with a high performance receiver at 
this frequency.  Then Parkes became one of the first telescopes to 
acquire this capability - using the 6.6-GHz receiver that had recently 
been purpose-built for methanol studies.  It allowed the discovery of 
many new 6035-MHz masers accompanying 1665-MHz masers (Caswell \& Vaile 
1995), sparking new interest in the transition.  
Much rarer is the maser emission from OH in an even more highly 
excited state, at 13.4 GHz.  Again, this is a frequency not commonly 
covered by high performance receivers on large telescopes.  From 1970 
until 2002, only one definite maser was known at this 
transition.  Improvements to the Parkes dish surface in 2003 and a 
receiver of modest performance then provided a combination viable 
for renewed observational effort, closely following similar renewed 
efforts at Effelsberg.  Parkes was able to detect 8 masers at this 
transition (Caswell 2004), most of them new and visible only from the 
southern hemisphere, and increasing the known number to 11, where the 
total still remains,  pending new Parkes observations, 

\section{Parkes and its role in spectroscopy with the LBA}

Dave Jauncey and John Reynolds will later describe Australian VLBI more 
fully, but here I must mention the role of the Australian LBA (Long 
Baseline Array) in maser studies.  

The LBA  baselines, from just the three ATNF elements (Parkes, the ATCA 
and Mopra), extend to 300km, very similar to MERLIN in the northern 
hemisphere, and allow us to do similar work, but in the richer fields of 
the southern Galaxy.

Spectroscopy with the LBA was still in the realm of pioneering work in 
1998 when we made our first observations of OH masers at 1665 and 1667 
MHz.  We  were able to observe both transitions 
with high spectral resolution in a single band covering a large velocity 
range, and simultaneously observing two polarizations.  These 
capabilities were able to solve  earlier problems of precise relative 
positional registration that had plagued earlier VLBI work.
The success of this ambitious project owed much to the ingenuity of John 
Reynolds who coped with each new problem as it arose.  The net result 
was a series of southern OH maser sites mapped at high resolution, and 
revealing their magnetic fields from the recognition in the spot 
distribution of Left and Right hand circular polarization of multiple 
Zeeman pairs (e.g. Caswell \& Reynolds 2001 and subsequent papers).  

As remarked earlier, OH maser studies of star formation regions 
mostly focus on the ground-state transitions at 1665 and 1667 MHz, 
but the 6035 and 6030-MHz excited-state transitions are even more  
valuable in clearly displaying the Zeeman pairs from which magnetic 
fields can be inferred.  A high performance 6-GHz receiver at Parkes was 
the catalyst to extend LBA observations to this 
transition.   In this case, it also allowed precise registration of 
maser spot distributions at 6035 MHz with those of the weaker 6030-MHz 
transition, in turn identifying Zeeman pairs and magnetic field 
distributions (Caswell, Kramer \& Reynolds 2011).  

The success of the LBA in this OH spectroscopy owes much to the large 
collecting area and high sensitivity provided by Parkes as a key 
element.

\section{The future - Galactic spiral structure, velocity field, and 
magnetic field}

What does the future look like for Galactic structure, high mass stars 
and masers?

Preliminary studies of the maser spatial and velocity distribution in 
the inner Galaxy are already at a stage where they can guide improved  
Galactic dynamics modelling, since current models are unable to account 
for the observations (Green et al. 2011),  but these are only the 
beginning of a much greater revolution.  

A landmark was achieved in 2006, with a demonstration that VLBI had 
matured to permit accuracies of better than 0.01 mas (Xu et al. 2006),  
allowing astrometric parallaxes and precise distance measurements to 
masers at the Galactic Centre and beyond, extending to the outer edge of 
the Galaxy (Reid et al. 2009).  
This achievement with the US VLBA at 12 GHz was shortly matched by 
similar measurements for 22-GHz water masers (which often accompany 
methanol masers) using the Japanese array VERA (Honma et al. 2007), and 
measurements of 6.6-GHz methanol masers using the EVN (Rygl et al. 
2010).  

So astrometry of masers can now provide a remarkable opportunity to map 
our Galaxy in detail, to reveal for the first time its precise geometry 
and velocity field.  These are the parameters that must be replicated 
by a valid dynamical model of the Galaxy. 
Southern and northern hemisphere telescopes will be needed to acquire 
the necessary observations and, in these endeavours, Parkes will be a 
key high sensitivity element in the southern LBA.  

Since OH masers are present at about half of the methanol sites, it will 
eventually be possible to associate a characteristic magnetic field at 
each site using Zeeman splitting, and thereby map the magnetic field of 
the Galaxy, with `in situ' measurements at each site, rather than the 
line-of-sight average fields that are commonly obtained by Faraday 
rotation measurements.

\section{Afterword}

In 1967, the outcomes of a conference held at Charlottesville on the 
topic   `Interstellar ionised hydrogen', were summarised by Gart 
Westerhout (Westerhout 1968).  At that time, the recently detected OH 
masers were the only known species of astrophysical maser.  The role of 
the masers  was uncertain.  In Gart's words:

`..how relevant are the OH (maser) observations to astrophysics....? 
Could it be that 
the emission is a pointer to regions of incipient star formation?   
Personally, I would say that the OH (maser) study is an extremely 
interesting intellectual exercise, which should be vigorously pursued, 
because such exercises lead almost always to completely new 
developments, and completely new ideas in both theories and techniques.  
But I don't think that the OH problem will contribute very much to our 
further understanding of the interstellar medium at large.'  

Has the pursuit of masers over the past 45 years been worthwhile?  
Perhaps the most emphatic answer is given by the title of a workshop at 
MPI, Bonn two years ago:  `Masers:  the ultimate astrophysical tools'!




\acknowledgments

I thank the conference organisers for the opportunity of contributing 
to this meeting, and expressing my gratitude to many close colleagues 
who have worked with me in the 40 years that I have enjoyed using the 
Parkes telescope.

\end{document}